\documentstyle[12pt,psfig]{article}
\def\reference{\parskip 0pt\par\noindent\hangindent 0.5 truecm}

\begin{document}
%
% Title
% Capitalise the title normally - do not use ALL CAPS.
%
\title{Annual Modulation in the Variability Properties of the IDV-Source 0917+624 ?}
%
% Authors
% Here comes the author(s) of the paper. Please add the appropriate author
% names for your paper and indicate within the $^...$ the number(s)
% which corresponds to the institute(s) of each author. In this example
% the second author has two institutional affiliations.
% Add or remove authors as required.
% **** IMPORTANT: Leave the closing curly bracket line as is. ******
\author{L. Fuhrmann$^{1}$,\and 
 T.P. Krichbaum$^{1}$, 
 G. Cim\`o$^{1}$, 
 T. Beckert$^{1}$,
 A. Kraus$^{1}$,\and
 A. Witzel$^{1}$,
 J.A. Zensus$^{1}$,
 S.J.Qian$^{1,2}$,
 B.J. Rickett$^{3}$
} % IMPORTANT: leave this curly bracket as the first character of this line.
% Date - leave this blank.
\date{}
\maketitle
% Institutions
% Here fill in your institute name(s) and address(es)
% The number in $^...$ indicates the author number.  For example
{\center
$^1$ Max-Planck-Institut f\"ur Radioastronomie, Bonn\\[3mm]
$^2$ Beijing Astronomical Observatory, China\\[3mm]
$^3$ University of California, San Diego\\[3mm]
}
% Abstract
% Simply place your abstract between the \begin{abstract} and
% \end{abstract} commands.
%
\begin{abstract}
\noindent
\small{There is new evidence which identifies seasonal changes of the variability time scale
in IDV-sources with refractive interstellar scintillation effects. Such a RISS-model takes the 
annual change of the Earth velocity relative to the scattering medium into account.
In September 1998 we found a remarkable prolongation of the variability time scale 
in the IDV-source 0917+624 with only small variations in flux density during a period of 5 days. 
This was explained as a seasonal effect, in which the velocity vector of the Earth and the 
interstellar medium nearly canceled. In order to further investigate the applicability of the model 
for 0917+624, we performed an Effelsberg 6cm-flux monitoring program over the course of one 
year. Since September 2000, the source appears to be remarkably inactive and yet (May 2001), 
no return to its normal, faster and stronger variability pattern is observed. Here, our
observational results and a possible explanation for the current quiescence are presented.} 
\end{abstract}
{\bf Keywords:}
% Place keywords here. Please write all keywords in lower case. PASA uses the
 %standard list of subject 
% headings adopted by The Astrophysical Journal and available from URL:
%   http://www.journals.uchicago.edu/ApJ/keywords_text.html
quasars: individual (0917+62), radio continuum: ISM, scattering
% A formatting command to add space between the author list and the body
% of the paper when printed. This spacing may be changed as desired.
%\bigskip
%
% Body of paper
%
\section{Introduction}
\begin{figure}
%\begin{center}
\hspace{2.7cm}
%\vspace{-0.5cm}
\psfig{file=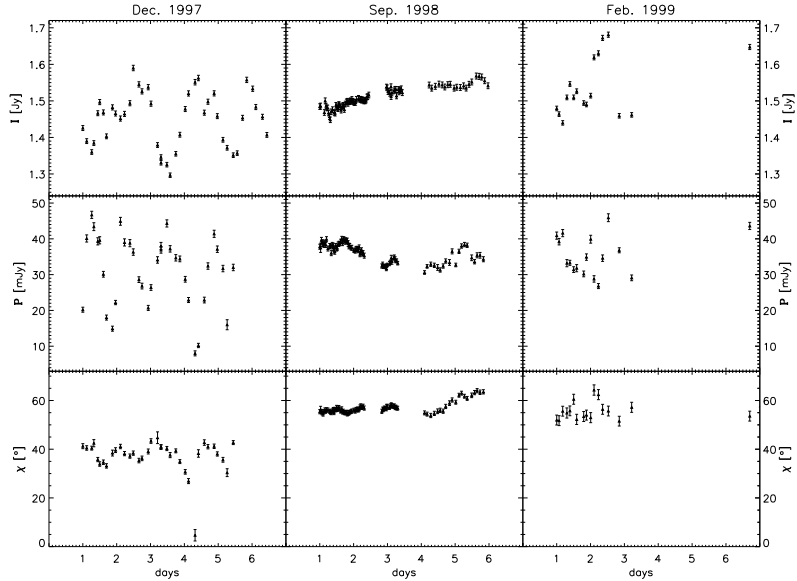,height=7cm,angle=90}
\caption{Observations of 0917+624 by Kraus et al. (1999) between December 1997 and February 1999. From top to bottom: 
total flux density I, polarized flux density P and polarization angle $\chi$ against time. Note the remarkable 
'slow down' in September 1998 and the return to fast variability in February 1999.}
\label{sep98}            
%\end{center}
\end{figure}

The quasar 0917+624 was one of the sources in the original IDV-sample in 1985, which showed 
variability on time scales shorter than one day (Intraday Variability, IDV, 
Heeschen et al. 1987). Since then, the source was regularly monitored over the last 15 years and 
generally was found to be strongly variable with amplitude variations on a $10-15\%$ level and on time scales in the 
range of 0.8-1.6 days. Furthermore, faster variability is detected in the polarized flux and 
the polarization angle. In polarization, larger amplitude variations (up to a factor of 2) are common and 
once a $180^{\circ}$-swing in the polarization angle could be observed (Quirrenbach et al. 1989). 
In September 1998, however, Kraus et al. (1999) detected a remarkable change in the 
variability properties: the rapid variability stopped and only a slow increase of $7\%$ in flux 
over the 5 day observing session has been observed (Fig. \ref{sep98}). Observations in February 1999 
then showed 0917+624 to be variable again on a timescale of 1.3 days. Kraus et al. (1999)
interpreted the slow-down in September as probably due to the ejection of a new jet component.
This temporarily leads to a source size larger than the scattering size (set by the ISM) and a 
reduction of the observed refractive interstellar scintillation (quenched RISS) (Rickett et al. 1995).
An alternative possibility to explain the behavior of 0917+624 within an ISS-model was given by 
Rickett et al. 2001 (see also Jauncey and Macquart 2001). They suggested that the observed change in 
the variability time scale depends on the time of the year and thus reflects the orbital motion of the 
earth relative to the ISM. Hence, using an ISS-model for 0917+624 as suggested by Rickett et al. (1995)
and taking the earth orbital motion around the sun with respect to the scattering medium into account
leads to a predicted slow-down in September and could explain the prolongation in September 1998. 
Since we have only one data point during this highly interesting time period, we performed a dedicated flux density 
monitoring program for 0917+624 over the course of 1 year, in order to test the annual modulation of IDV
in this source (and two others).
%%%%%%%%%%%%%%%%%%%%%%%%%%%%%%%%%%%%%%%%
%%%%%%%%%%%%%%%%%%%%%%%%%%%%%%%%%%%%%%%%%%%
% Place contents of first section here.
\section{Annual Modulations in IDV}
\begin{center}
\begin{figure}
%\begin{center}
%\hspace{2.7cm}
\psfig{file=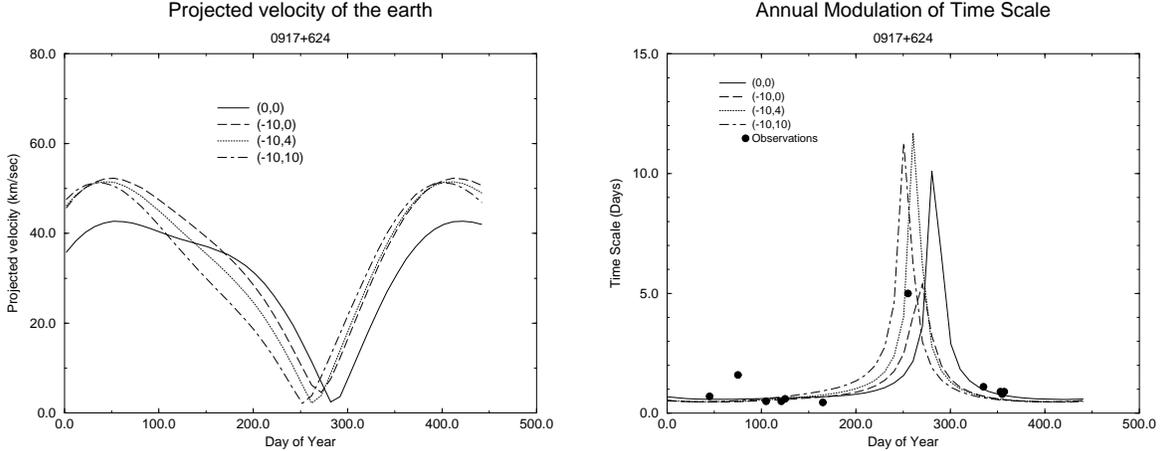,height=6cm,angle=-90}
\caption{Left: the velocity of the earth relative to the LSR projected perpendicular to the direction of 0917+624 versus day of the year
for 4 different velocities of the screen. Note the strong decrease near day 285. Right: resulting 'slow-down-peak' in the predictions of
the variability time scale. We overplot the predicted and the observed time scales and find a good agreement (see text).}
\label{mod}            
%\end{center}
\end{figure}
\end{center}

Recent evidence for a seasonal dependence of the variability timescale comes from the highly variable 
quasar J1819+3845 (Dennett-Thorpe and de Bruyn 2000). Westerbork observations during a period of 12 month 
indicate a variation of the IDV time scale by a factor of $\sim10$ and a clear 'slow-down-peak' near day 280.  
In the following year the prolongation could nicely be reproduced (de Bruyn, priv. com.).
However, in addition to the earth motion with respect to the Local Standard of Rest (LSR), a
medium velocity of $\sim 25~kms^{-1}$ is required.\\
In the standard ISS-model, the observer moves through a spatial scintillation pattern caused by a screen of scattering 
material at a distance D. The variability time scale is then determined by two quantities: the spatial scale
of the pattern and the observers motion projected onto the scintillation pattern. In order to explain 
the annual variation of the IDV time scale in terms of ISS, the relative motion of the Earth with respect to the 
scattering medium is an important parameter and has to be taken into account. This relative motion is composed of 
three different velocities: i) the Earth orbital motion around the Sun; ii) the Sun's motion towards the solar apex; 
iii) the velocity of the medium with respect to the LSR. In Fig. \ref{mod} the velocity of the Earth relative
to the LSR projected perpendicular to the direction of 0917+624 is displayed versus day of the year. Near day number 285
the velocity of the Earth's orbital motion nearly cancels versus the projected velocity of the sun relative to the LSR.
This effect leads to a seasonal dependence in the variability time scale: while the predicted time scale should 
remain nearly constant over a large fraction of the year, a substantial prolongation in September and October (days 250-330) 
should occur. In Fig. \ref{mod} we overplot the predicted and observed time scales and we find a general 
agreement between the observations and the predictions. The data point in September from Kraus et al. (1999) lies 
in the period of the predicted increase. The other data points come from our observations performed during 1989 - 2000.    
%%%%%%%%%%%%%%%%%%%%%%%%%%%%%%%%%%%%%%%%%%%%%%%%
%%%%%%%%%%%%%%%%%%%%%%%%%%%%%%%%%%%%%%%%%%%%%%%%
% Place contents of next section here.
\vspace{-1cm}
\section{Observations and data reduction}
In order to investigate the predictions of the above model for 0917+624 in more detail, we started a monitoring
program at 6cm wavelength over the last year. The observations were performed from 
September 2000 to May 2001 at the 100m telescope of the MPIfR in Effelsberg. During this time 
22 single observing sessions with a duration of a few hours up to 3 days were performed.
In this experiment 0917+624 was one source within a small sample of IDV-sources and has been observed 
with a dense duty cycle of about 2 measurements every 1.5 hrs. In addition, two non-IDV sources, 0951+699
and 0836+73, were included as secondary calibrators. We observed standard flux density calibrators 
like 3C286, 3C295, 3C48 and NGC7027 frequently. Since all measured sources are point-like and sufficiently 
strong at 6cm ($\geq 0.5$ Jy) we were able to measure the flux densities with cross-scans in azimuth and elevation 
(eg. Quirrenbach et al. 
1992, Kraus 1997). We used the standard IDV data reduction procedure of the Effelsberg telescope
(for details see Quirrenbach et al. 1992 and Kraus 1997). Our secondary calibrators were observed with similar
duty cycles as the program sources to allow for a systematic elevation- and time-dependent correction of the measured 
flux densities. The resulting 
flux density errors are composed of the statistical errors from the reduction process (Gaussian fits of averaged sub-scans)
and a contribution from the systematics
as seen in
our secondary calibrators (weather). Our final errors lie in the range of $\le0.5-1\%$.\\ 
Since a polarimeter is integrated in the 6cm receiver, we also obtained full polarization information. The reduction 
and analysis of the polarization data is still in progress and will be discussed elsewhere 
(Fuhrmann et al. in prep.).
%%%%%%%%%%%%%%%%%%%%%%%%%%%%%%%%%%%%%%
\begin{center}
\begin{figure}[t]
%\begin{center}
%\hspace{2.7cm}
\psfig{file=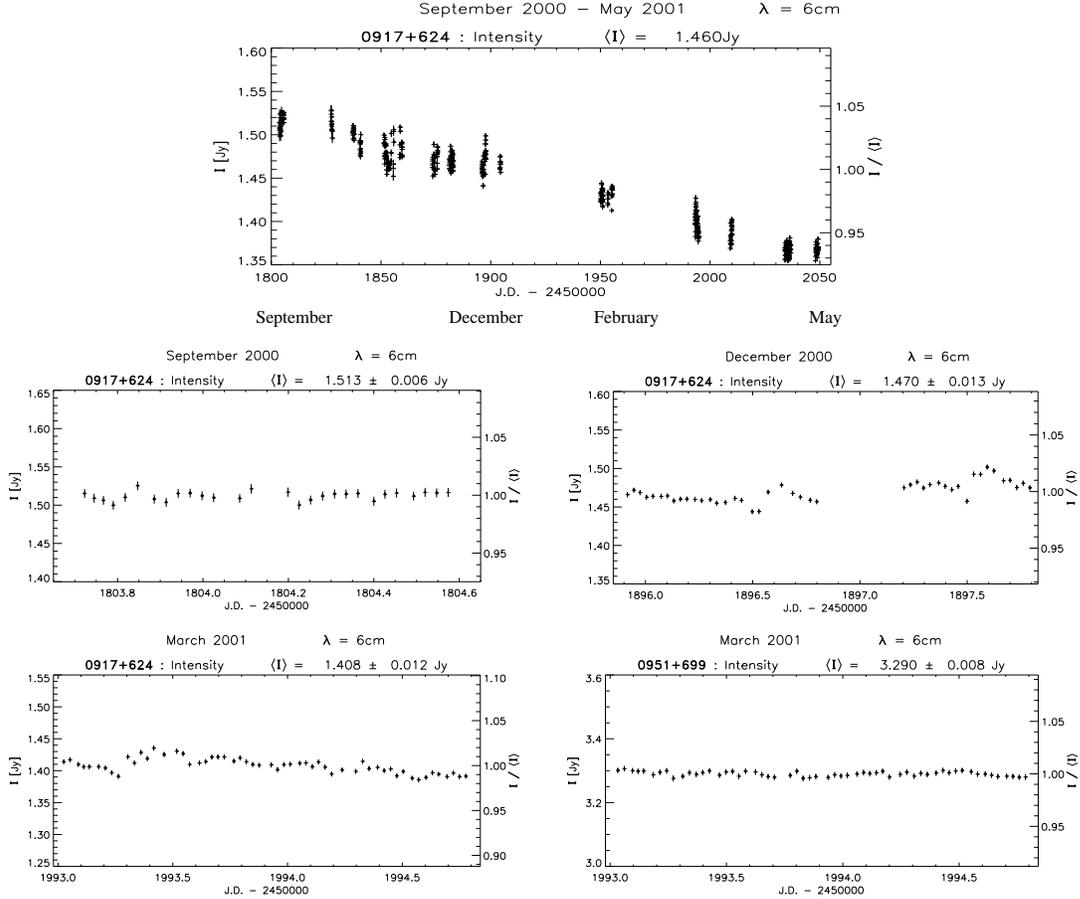,height=12cm,angle=0}
\caption{Summary of our observations between September 2000 and May 2001: at the top, the longterm light curve of 0917+624
with a trend of a $\sim10\%$-decrease is displayed. In addition, three individual light curves from September 2000 ($m=0.4\%$), 
December 2000 ($m=0.9\%$) and March 2001 ($m=0.8\%$) are plotted. For comparison, the lower right panel shows one of our secondary 
calibrators in March 2001 ($m=0.2\%$).} 
\label{lc}            
%\end{center}
\end{figure}
\end{center}
%%%%%%%%%%%%%%%%%%%%%%%%%%%%%%%%%%%%%   
\vspace{-1.5cm}
\section{Results}
In Fig. \ref{lc} we summarize our results. Since September 2000, all 22 individual light curves show a very inactive 
source with only few percent amplitude variations. At the top of Fig. \ref{lc} the longterm light curve of 0917+624 
over the full monitoring period of 9 month is displayed. During this time interval, we detected a 
longterm trend with a monotonic, $\sim10\%$-decrease in total flux since October 2000. In addition, Fig. \ref{lc} 
shows 3 individual light curves measured in September 2000, 
December 2000 and March 2001. For comparison, the March 2000 light curve of one of our secondary calibrators 
(0951+699) is displayed. All curves of 0917+699 show only very low amplitude variations of about 2-3\% ("ripples")
on very short time scales of $\sim0.3-0.4$~days. Since September, no significant IDV is seen in 0917+624 
and the strong, pronounced IDV seen at earlier times is completely missing.
\vspace{-0.3cm}
\section{Discussion}
\begin{figure}
\vspace{-1cm}
\begin{center}
\hspace{0.3cm}
\psfig{file=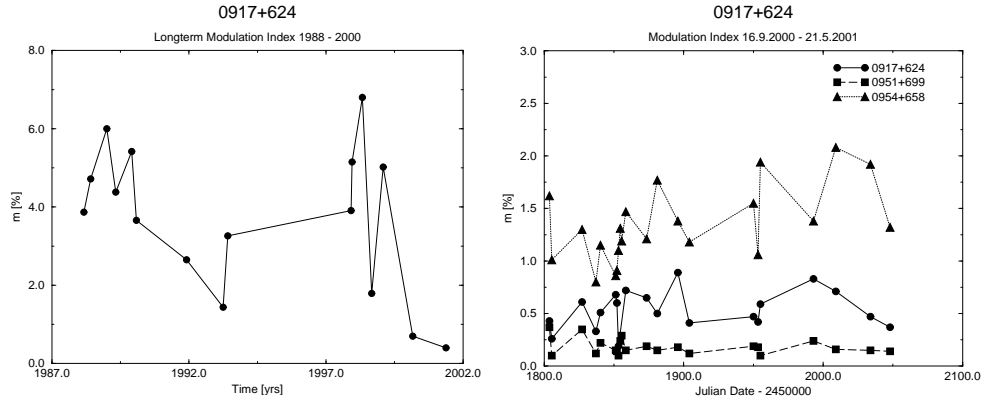,height=5.3cm,angle=-90}
\caption{Left: longterm modulation index distribution between 1988 and 2000 representing the strong variability 
of 0917+624 up to year 2000. Right: modulation index during our monitoring campaign 2000/2001. In addition, the IDV-source 
0954+658 and the secondary calibrator 0951+699 are displayed.}
\label{m}            
\end{center}
\vspace{-0.4cm}
\end{figure}

It can be seen from Fig. \ref{lc}, that 0917+624 passed through a dramatic change of its variability properties when compared
to the past. The change can clearly be seen by plotting a light curve of our new campaign on the same scale with 
data from earlier observations. This is realized in Fig. \ref{comp}: while in June 1993 0917+624 showed its normal,
strong variability pattern with $\sim15\%$-amplitude variations on a time scale of $\sim0.5$ days, the variations in March 2001 appear 
dramatically low. In order to compare the strength of the variability, we determined the modulation index $m=100\times\sigma_{S}/<S>$
where $\sigma_{S}$ is the rms flux density variation. It can be seen from Fig. \ref{m} that during our recent monitoring
m has been always
low: $m\le0.9$. In the most extreme case, this is a factor $\sim7$ lower than the variations observed in the past.
In order to explain the behavior of 0917+624 in the context of RISS and the annual modulation effect, however, we see 
three main problems: i) Other IDV sources in the same region of the sky (eg. 0716+71, 0954+65) do not show the expected 
seasonal dependence of the variability time scale. ii) The orbital motion of the earth should affect the time scale, 
but not the modulation index m (Rickett et al. 2001), which also appears to be variable (Fig. \ref{m}). iii) Our monitoring 
program presented here has not yet revealed a restart of the strong and pronounced variability in 0917+624 since 
September 2000. This contradicts the model predictions of Rickett et al. (2001), which claimed recovery 
to the "normal" variability already at the end of last year (see Fig. \ref{mod}).\\ 
We suggest two possible scenarios in order to explain the present anomalous behavior of 0917+624: 
Firstly, the interstellar medium is far more complex than presently thought and a change in the scattering properties 
of the screen might explain the change of the variability mode in 0917+624. If the spatial scale of the scintillation pattern 
decreased, then the source size could quench the ISS. For weak ISS or RISS this would imply either a decrease in the strength of 
turbulence in small, localized regions of the ISM or an increase of the distance to such a region. 
%A change of the angular scattering size, implying either a
%decrease in the strength of turbulence in small regions of the ISM or an increase of the distance to such a region
%might explain the change of the variability mode in 0917+62. A possible scenario could invoke a 
In turn this would require a very inhomogeneous ISM, with moving `clouds' or `layers' at different distances. 
The inhomogeneities then have to occur on angular scales of less than the angular separation from 0917+62 to other 
sources ($\leq 5-10^{\circ}$), the latter still showing pronounced IDV. In this context, the present situation of nearly 
no variations in 0917+624 could be explained by either a ``hole'' in the screen or a moving foreground layer, which was 
present during the last 15 years and has suddenly disappeared within the last year.\\ 
Secondly, an alternative interpretation could be based on intrinsic variations in the VLBI structure of the 
source between 1998 and 2000. The dominant components, which in the past were responsible for RISS, either disappeared or increased
their intrinsic size, hence producing quenched scattering. Recent evidence for such effect is coming from VLBI observations 
at 15 GHz (see T.P. Krichbaum, this conference). Multi-epoch observations during 1999 and 2000 reveal structural expansion 
in the central region with 5-6c. Extrapolating the motion of one jet component backwards in time leads to a time of ejection 
shortly before the date, when 0917+624 ceased varying for the first time (September 1998, Kraus et al. 1999). It is therefore 
very tempting to assume that the present quiescence in the variability of 0917+624 once more is caused by blending effects of 
the inner, scintillating region by a another new component. This component probably was ejected during the first half of the year 
2000 and should become visible in future VLBI observations.\\
We conclude, that the annual modulation effect in 0917+624 may at present not be visible due to structural changes in the central 
region and hence, only strongly quenched 
%%%%%%%%%%%%%%%%%%%%%%%%%%%%%%%%%%%%%%
\begin{center}
\vspace{-0.5cm}
\begin{figure}[b]
%\begin{center}
%\hspace{2.7cm}
\psfig{file=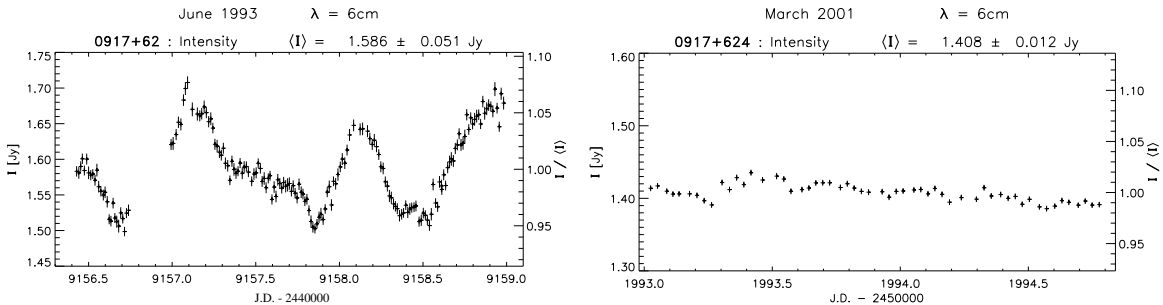,height=4cm,angle=-90}
\caption{Comparison of two periods with different variability pattern: the left curve shows 0917+624 in June 1993
with its normal, strong variability pattern ($m=3.3\%$). On the right, a recent light curve from March 2001 (plotted on the same scale) 
is displayed ($m=0.8\%$). Note the low variations compared to the past.}
\label{comp}            
%\end{center}
\end{figure}
\end{center}
%%%%%%%%%%%%%%%%%%%%%%%%%%%%%%%%%%%%%
\newpage
\noindent
scattering could be observed during the last 9 months. In order to detect a seasonal 
dependence of the variability time scale, less complex and more point-like sources (eg. J1819+3845) would provide a better 
opportunity to investigate the annual modulation model.                   
%%%%%%%%%%%%%%%%%%%%%%%%%%%%%%%%%%%%%
%%%%%%%%%%%%%%%%%%%%%%%%%%%%%%%%%%%%%
%
% Add as many section titles/contents as required.
%
% If you have subsections then use the
% \subsection{Subsection Title}
% command and if you have subsubsections then use the
% \subsubsection{Subsubsection Title}
% command.  To use these commands, 
% first remove the % from the start of the line.
% It is preferable to embed your figures in the text. 
% One way to do this is to use the psfig style file and use the following
% commands to include the figures:
% \begin{figure
% \begin{center}
% \psfig{file=filename.ps,height=10cm}
% \caption{Write your figure caption here.}
% \label{figlabel}            % for cross-references
% \end{center}
% \end{figure}
% To use the above commands, first remove the % from the beginning of
% the lines and then fill in your own values etc as appropriate.
% Tables
% Please consult previous issues of PASA
%  to see how tables are to be formatted.
% Place acknowledgments here. Omit above \section command if there
% are no acknowledgments.
\section*{References}
% PASA uses the same conventions as ApJ for journal abbreviations.  Sample
% references are as follows. 
% Please follow the same format for your references.
%\reference Author, A. B., Anotherauthor, C. D. \and Thirdauthor, E. F. 1990, 
% PASA, 7, 350
% for a journal article, or
% \reference Author, A.B. \and Anotherauthor, C. D. 1990, in This is a Book %Title, ed. C. D. Editor, (City: Publisher Name), 437
% for a book.
\reference Dennett-Thorpe, J. \& de Bruyne, A.G., 2000, ApJ, 529, L65
\reference Heeschen, D.S., et al. 1987, AJ, 94, 1493
\reference Jauncey, D.L. \& Macquart, J.-P., AA, 370, L9  
\reference Kraus, A., et al. 1999, AA, 352, L107
\reference Kraus, A., 1997, Ph.D. thesis, University of Bonn, Germany
\reference Quirrenbach, A., et al. 1989, AA, 226, L1
\reference Quirrenbach, A., et al. 1992, AA, 258, 279
\reference Rickett, B.J., et al., 2001, ApJ, 550, L11
\reference Rickett, B.J., et al., 1995, AA, 293, L479
% Add as many references as required.
\end{document}